\documentstyle[emulateapj,psfig,apjfonts]{article}
\def\gsim{\mathrel{\raise0.35ex\hbox{$\scriptstyle >$}\kern-0.6em
\lower0.40ex\hbox{{$\scriptstyle \sim$}}}}
\def\lsim{\mathrel{\raise0.35ex\hbox{$\scriptstyle <$}\kern-0.6em
\lower0.40ex\hbox{{$\scriptstyle \sim$}}}}

\submitted{Accepted for publication in The Astrophysical Journal Letters}

\lefthead{Kodama et al.}
\righthead{Environmental Variation of Cluster Galaxy Colors at $z=0.41$}

\begin{document}

\title{The Transformation of Galaxies within 
the Large Scale Structure around a $z=0.41$ Cluster \altaffilmark{1}}

\author{T.\ Kodama\altaffilmark{2}, Ian Smail\altaffilmark{3},
F.\ Nakata\altaffilmark{2}, S.\ Okamura\altaffilmark{2},
R.\ G.\ Bower\altaffilmark{3}
}
\altaffiltext{1}{Based on data collected at 
Subaru Telescope, which is operated by 
the National Astronomical Observatory of Japan.}
\altaffiltext{2}{Department of Astronomy, University of Tokyo, Hongo,
Bunkyo-ku, Tokyo 113--0033, Japan}
\altaffiltext{3}{Department of Physics, University of Durham, South 
Road, Durham DH1 3LE, UK}

\setcounter{footnote}{3}

\begin{abstract}
We present deep, panoramic multi-color imaging of the distant rich
cluster A\,851 (Cl\,0939+4713, $z=0.41$) using Suprime-Cam on Subaru.
These images cover a 27$'$ field of view, $\sim 11 h_{50}^{-1}$\,Mpc at
$z=0.41$, and by exploiting photometric redshifts estimated from our
$BV\! RI$ imaging we can isolate galaxies in a narrow redshift slice at
the cluster redshift.  Using a sample of $\sim 2700$ probable cluster
members brighter than $0.02 L^\ast_V$, we trace the network of
filaments and subclumps around the cluster core. The depth of our
observations, combined with the identification of filamentary
structure, gives us an unprecedented opportunity to test the influence
of the environment on the properties of low luminosity galaxies.  We
find an abrupt change in the colors of $\lsim 0.1L_V^\ast$ galaxies at
a local density of 100 gal.\,Mpc$^{-2}$, with the population in lower
density regions being predominantly blue, while those in higher density
regions are red.  The transition in the color-local density behavior
occurs at densities corresponding to subclumps within the filaments
surrounding the cluster. Identifying the sites where the transition
occurs brings us much closer to understanding the mechanisms which are
responsible for establishing the present-day relationship between
environment and galaxy characteristics.
\end{abstract}


\section{Introduction}

Clusters of galaxies are a very visible constituent of the structure of
the Universe at the present-day.  Redshift surveys of the local Universe,
$z<0.1$, illustrate the filamentary nature of large scale structure (e.g.\
Peacock et al.\ 2001), and the presence of clusters at the intersections
of these filaments and walls.  N-body simulations have been particularly
successful in reproducing these filamentary features, showing that they
are a natural consequence of gravitationally-driven structure formation
(e.g.\ Moore et al.\ 1998).  The precursors of the filaments should be
present around distant clusters, containing many of the galaxies which
will subsequently infall onto the virialized core and form the cluster
population we see today.

The influence of environment on the star formation histories of
the galaxies is a critical question for models of galaxy evolution.
The striking variation in the stellar populations of galaxies in different
environments (e.g.\ Larson, Tinsley \& Caldwell 1980; Butcher \& Oemler
1984; Balogh et al.\ 1999; Norberg et al.\ 2001) clearly indicates
the importance of such environmental influences on star formation.
Since clusters (which are dominated by passive galaxy populations)
are continuously growing through the accretion of galaxies  and groups
from the field, which is dominated by actively star-forming galaxies,
this activity must be quenched during the assimilation of the galaxies
into the cluster.  This transformation is a key process in creating the
environmental dependence of galaxy properties, and may also underpin the
observed evolution of galaxy properties in distant clusters (e.g.\ Butcher
\& Oemler 1984; Kodama \& Bower 2001). However, the  physical mechanism
which is responsible for these changes has not yet been identified (Moore
et al.\ 1996; Abadi et al.\ 1999; Balogh et al.\ 2000).  Recent studies
have begun to focus on tracing the variation of galaxy properties from
the cores of clusters, out to the surrounding field in an attempt to
identify the environment where the decline in the star formation  in
accreted galaxies begins (e.g.\ Abraham, et al.\ 1996; Balogh et al.\
1999; Pimbblet et al.\ 2001).

The advent of Suprime-Cam, a revolutionary wide-field camera on the Subaru
telescope, has provided an unique new tool to tackle programs requiring
deep, high quality imaging across large fields.  In this {\it Letter},
we analyze unique deep, panoramic, multi-color imaging of the $z=0.41$
cluster A\,851.  We estimate photometric redshifts for galaxies across
the $27'$ field and isolate a narrow redshift slice around the cluster.
Our two-dimensional map covers $11\times 11$\,Mpc,
over a much wider field of view than the previous studies
(e.g.\ Iye et al. 2000), allowing us to identify
a wealth of filamentary structures extending from the cluster core.  We
use the depth and wide range in environment spanned by our observations,
over two orders of magnitude in galaxy surface density, to investigate
the dependence of faint  galaxy properties on local environment.

\S2 describes the observations and reduction, \S3 details the analysis
of our photometric catalog and  discusses our results, while \S4  gives
our main conclusions.  We adopt $H_o=50$\,km\,s$^{-1}$\,Mpc$^{-1}$
and a $q_o=0.1$ cosmology.

\section{Observations, Reduction and Analysis}
\subsection{Observations and Reduction}

We obtained deep, multi-band imaging of the rich cluster A\,851
(Cl\,0939+4713), using the optical mosaic camera, Suprime-Cam, on the
8.3-m Subaru Telescope, Mauna Kea, Hawaii.  These observations were
undertaken on 2001 January 21--22 as a common-user program during the
first semester of telescope operations.  The weather conditions were
good and photometric during the observations. Total exposure times
of 3.6, 2.2, 4.0 and 1.3\,ks were obtained in $BV\! RI$ respectively,
with seeing measured off the stacked frames of 1.1, 0.7, 1.0, 0.7$''$.
These frames have 5-$\sigma$ limiting magnitudes of $B=27.0$, $V=26.5$,
$R=26.2$ and $I=25.0$, sufficient to detect a passively evolving
early-type galaxy as faint as $0.02L^\ast_V$ at $z=0.41$ in all passbands.

We used {\sc iraf} and purpose-written software: {\sc nekosoft}
(Yagi 1998); developed by the Suprime-Cam team, to reduce the data.
This entailed bias subtraction, flat fielding using  super sky-flats
constructed from the median of the dithered science frames ($>$9
frames), matching the point spread function (PSF) between the mosaic
chips, relative calibration of the fluxes between chips, mosaicing,
and photometric calibration using standard stars from Landolt (1992).
We then constructed an $I$-band selected sample using SExtractor v.2.1.6
(Bertin \& Arnouts 1996).  For each object we use the {\sc phot} package
in {\sc iraf} to measure photometry within a 3$''$ diameter aperture
in all the passbands, after matching the PSF.  Hereafter, we employ
the aperture photometry for colors and SExtractor's  {\sc mag\_best}
for total magnitudes.  The final catalog used in our analysis has 15,055
galaxies brighter than $I=23.4$ ($M^\ast_V+4$ at $z=0.41$).

\subsection{Foreground/background subtraction}

To trace the properties of galaxies inhabiting the large scale
structure around the cluster core, we need to map the distribution of
cluster members out to very low density regions.  The key requirement
for our analysis is therefore the removal of unassociated galaxies in
the foreground and background, to maximize the contrast of the cluster
on the sky.

Spectroscopy is the ideal method to remove the field contamination from
our sample.  However, it is not practical to obtain spectroscopy for
the $\sim 15$,000 very faint galaxies required for our analysis.  We
therefore exploit photometric redshift techniques, as an observationally
efficient and reliable method to map the three dimensional distribution
and properties of faint galaxies over a large field.  We apply our
photometric redshift code (Kodama, Bell \& Bower 1999) to the galaxies
in our $BV\! RI$ catalog.  To test the reliability of our photometric
redshifts we compare the predicted redshifts with spectroscopic
measurements for the 67 confirmed cluster members from Dressler et al.\
(1999).  For S/N\,$>10$ detections, the dominant source of uncertainty
in the photometric redshifts is the match between the model spectrum
and the true galaxy spectrum. Thus, although the magnitude limit of our
photometric sample is fainter than that of the spectroscopic sample,
the error estimated from this comparison is a good indication of the
uncertainties for {\it all} galaxies with $I<23.4$.  The photometric
redshifts, $z_{\rm phot}$, for the spectroscopic members exhibits a tail
at lower redshifts, showing that they tend to be underestimated for some
of the  galaxies by up to $|\Delta z|\sim 0.1$. This is largely due to
the lack of $U$-band data, which is important to discriminate the blue
cluster members at $z=0.4$ from the foreground galaxies (e.g.\ Kodama
et al. 1999).  We note that this modest photometric bias will only serve
to weaken the trends we uncover in \S3.2.  We adopt a range of
$0.32<z_{\rm phot}<0.48$ for the photometric membership
to ensure that we include the bulk of the cluster population, $>80$\%,
while still reducing the field contamination by a factor of $\sim$10 at
$I=23.4$.

As a final step, we estimate the expected field contamination in our
photometric redshift slice using similar observations of a blank field.
This correction accounts both for field galaxies genuinely in the cluster
redshift slice and for galaxies whose redshifts are miss-classified by
the photometric analysis.  In this case we use comparably deep $BVRi'$
imaging of the `Subaru/XMM Deep Field'  over a similarly large area,
618\,arcmin$^2$ (Ouchi et al.\ 2001).  After transforming the SDSS
$i'$-band to Cousins $I$-band (Fukugita et al.\ 1995), we apply the same
photometric redshift code to the galaxies in this field observed in the
same combination of passbands $BV\! RI$ and adopt the same redshift cut
of $0.32<z_{\rm phot}<0.48$.  We find that the averaged surface number
density of field galaxies down to $I=23.4$ that fall within this redshift
range is $2.29\pm 0.06$ arcmin$^{-2}$.  We  subtract this remaining
contamination statistically  (\S3.2).

\section{Results and Discussion}

\subsection{Large scale structure}

Having applied our photometric redshift selection, it is straightforward
to map out the distribution of galaxies which are likely to be associated
with the cluster.  The full field of our observations is shown in Fig.~1.
Several large scale structures are visible around the cluster core:
two large subclumps $\sim 10'$ (4\,Mpc) to the West and South and a
number of filamentary extensions coming out directly from the core.
Importantly, most of these extensions from the core  are aligned with
the surrounding subclumps.  The alignment of structures visible to the
Northeast, Northwest and South are all good examples, each extending
out to 3--5\,Mpc.  Since the field contamination is only $<$20\% at the
lowest density contour, most of the structures traced by the contours
in Fig.~1 are expected to be real.

The structures identified in this region are qualitatively similar to
those seen in cosmological simulations of the growth of clusters which
exhibit the filamentary/clumpy substructures on similar scales (e.g.\
Ghigna et al.\ 1998).  It appears therefore that we are witnessing
A\,851 as it assembles through the accretion of galaxies and groups
along the filaments onto the cluster core from the surrounding field.
Many of the systems in these filaments will have been assimilated into
the cluster population by the present-day, given their likely infall
speeds of $\gsim 1000$\,km\,s$^{-1}$ ($\sim$1~Mpc/Gyr).

%
%
\centerline{\psfig{file=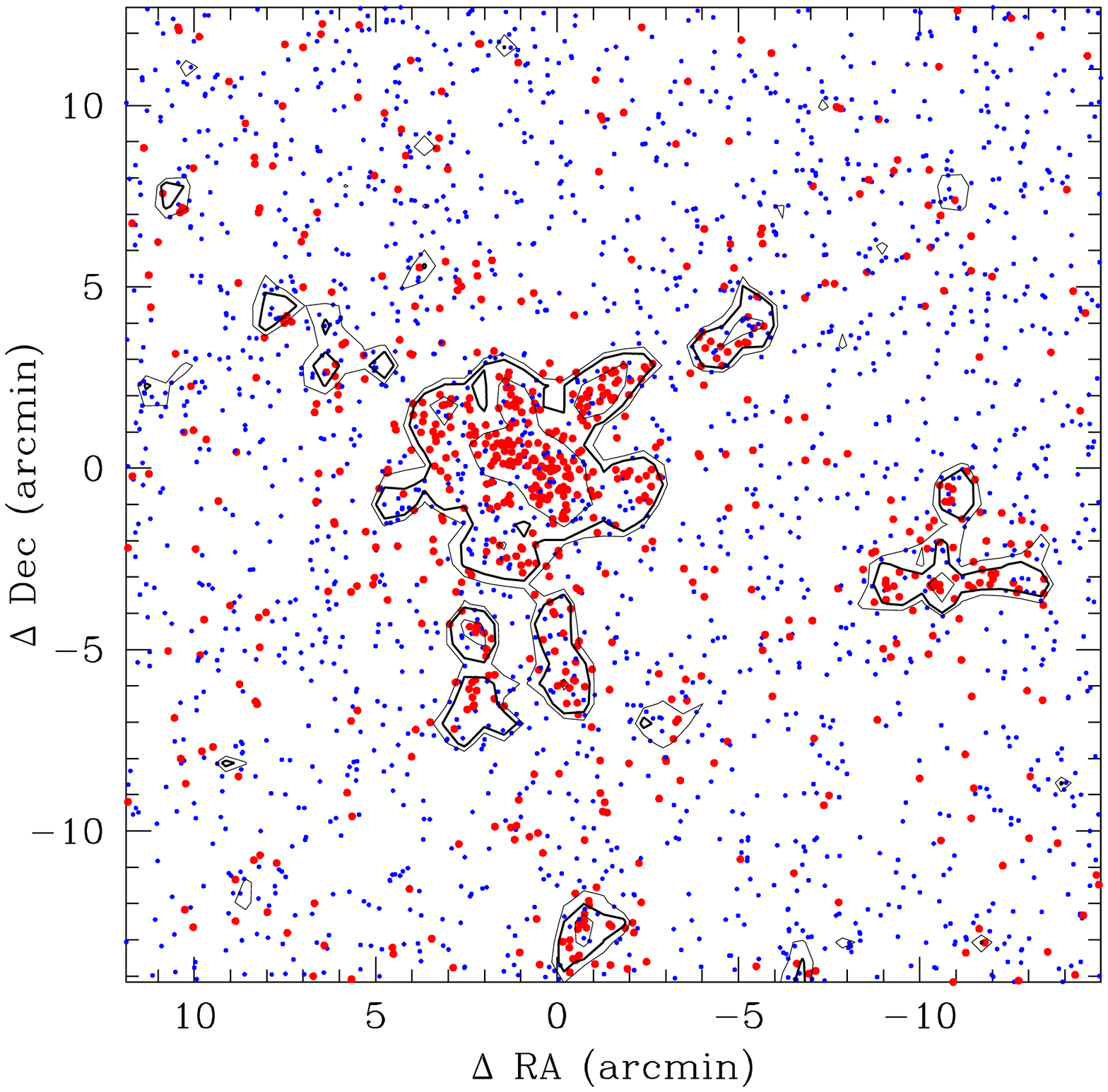,width=3.0in}}
\noindent{\scriptsize \addtolength{\baselineskip}{-3pt}
{\sc Fig.~1.} ---
A wide-field map of A\,851 from our Suprime-Cam $BV\! RI$ imaging.
The field is 11\,Mpc on a side.  The points identify cluster members based
on photometric redshifts.  Note that the remaining field contamination
is not subtracted (see \S3.2). The large and small circles identify the
color of galaxies redder or bluer than $(B-I)=3.4-0.1\times(I-17.0)$,
corresponding to 0.5 mag bluer than the color-magnitude sequence,
respectively.  The contours indicate the local surface density of
the member galaxies defined by 10 nearest neighbors, corresponding to
$\log_{10}\Sigma=1.8$, 2.0 (thick) and 2.5\,Mpc$^{-2}$ (after correcting
for field contamination; see \S3.2), respectively.  The thick contour
traces the boundary where the color distribution of galaxies changes
dramatically (see \S3.3).  Large scale structure is clearly visible
around the cluster core, with several filaments and subclumps coming
out from the main body of the cluster and extending to $\gsim$5\,Mpc.
The overall shape resembles an `octopus', with a round head and many legs.

}
\medskip

The combination of the depth of our observations and their complete
sampling, compared to spectroscopic surveys, means that our map of this
$z=0.41$ cluster provides one of the most detailed views of large scale
structure around clusters (West et al.\ 1995; Bardelli et al.\ 2001;
Connolly et al.\ 1996; Lubin et al.\ 2000; Clowe et al.\ 2000; Abraham
et al.\ 1996).  The application of the photometric redshift technique
to multi-color deep, panoramic CCD imaging allows us for the first time
to efficiently trace weak filamentary structures across large fields of
view. Furthermore, because the cluster is at relatively high redshift,
the contrast between the red galaxies in the cluster core and the blue
colors of the field galaxy population is greatly enhanced, providing
a much stronger gradient across the transistion zone between these two
regimes (\S3.2).

We have also compared the structures in Fig.~1 with the X-ray emission
from a 14.2-ks {\it ROSAT PSPC} exposure (Schindler et al.\ 1998).
However, we cannot see any significant emission from the subclumps or
filaments, except for the cluster core.  This is not surprising given the
steep slope of the mass-X-ray luminosity relation seen in local groups
(Mulchaey 2000).

\setcounter{figure}{1}
%
%
\begin{figure*}[htb]
\centerline{\psfig{file=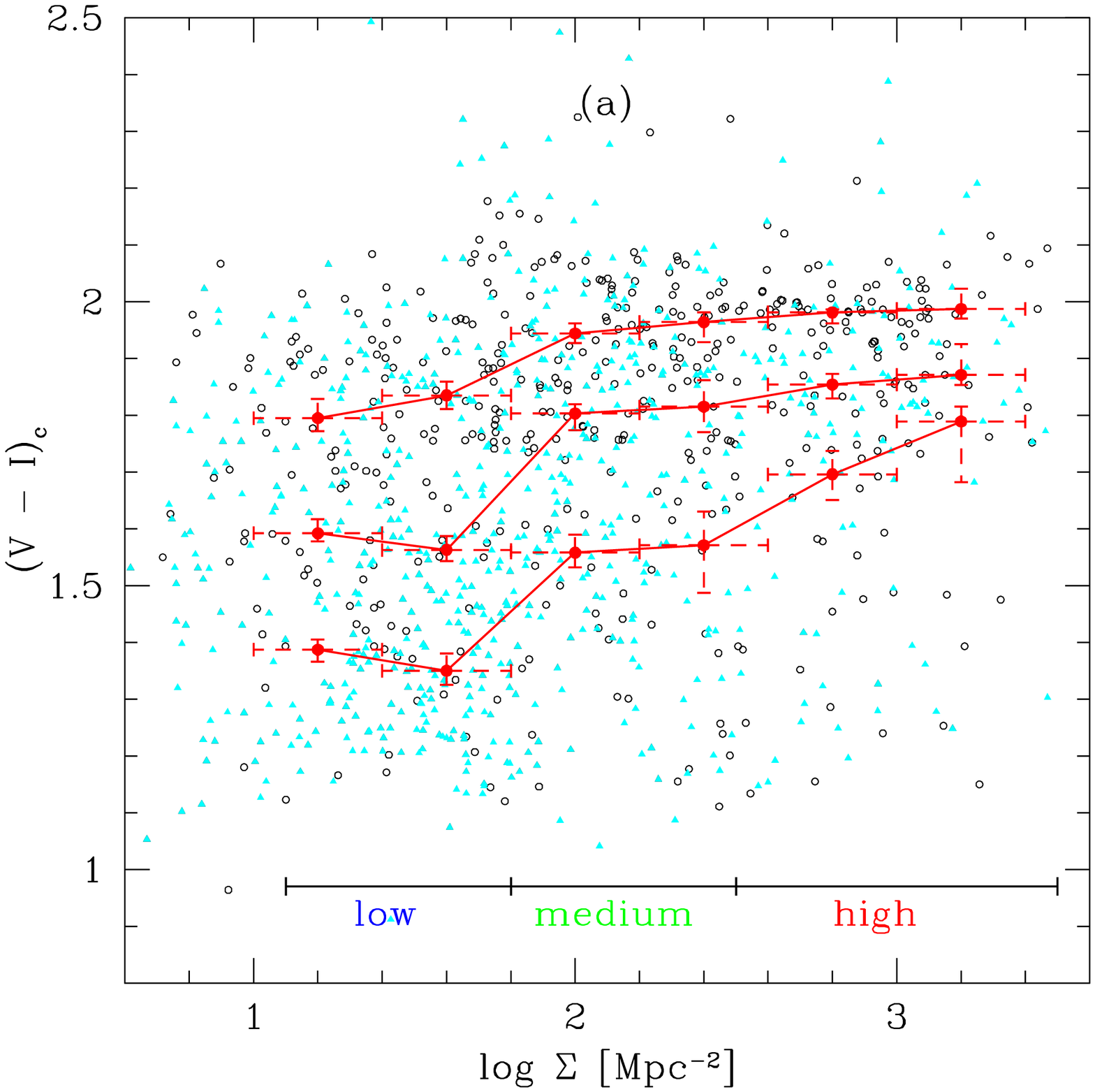,width=2.45in,angle=0}
\hspace*{0.0in}
\psfig{file=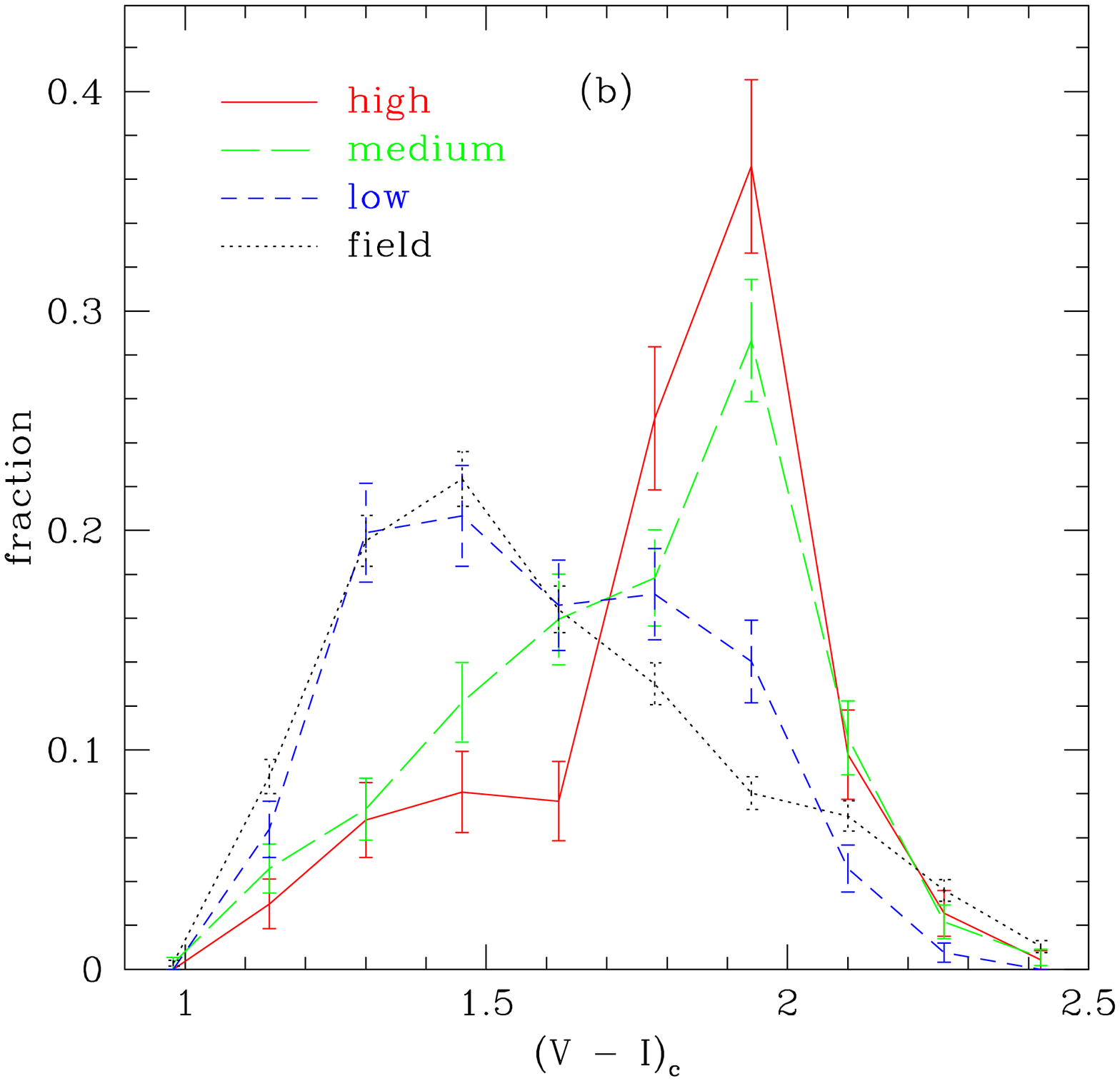,width=2.45in,angle=0}
\hspace*{0.0in}
\psfig{file=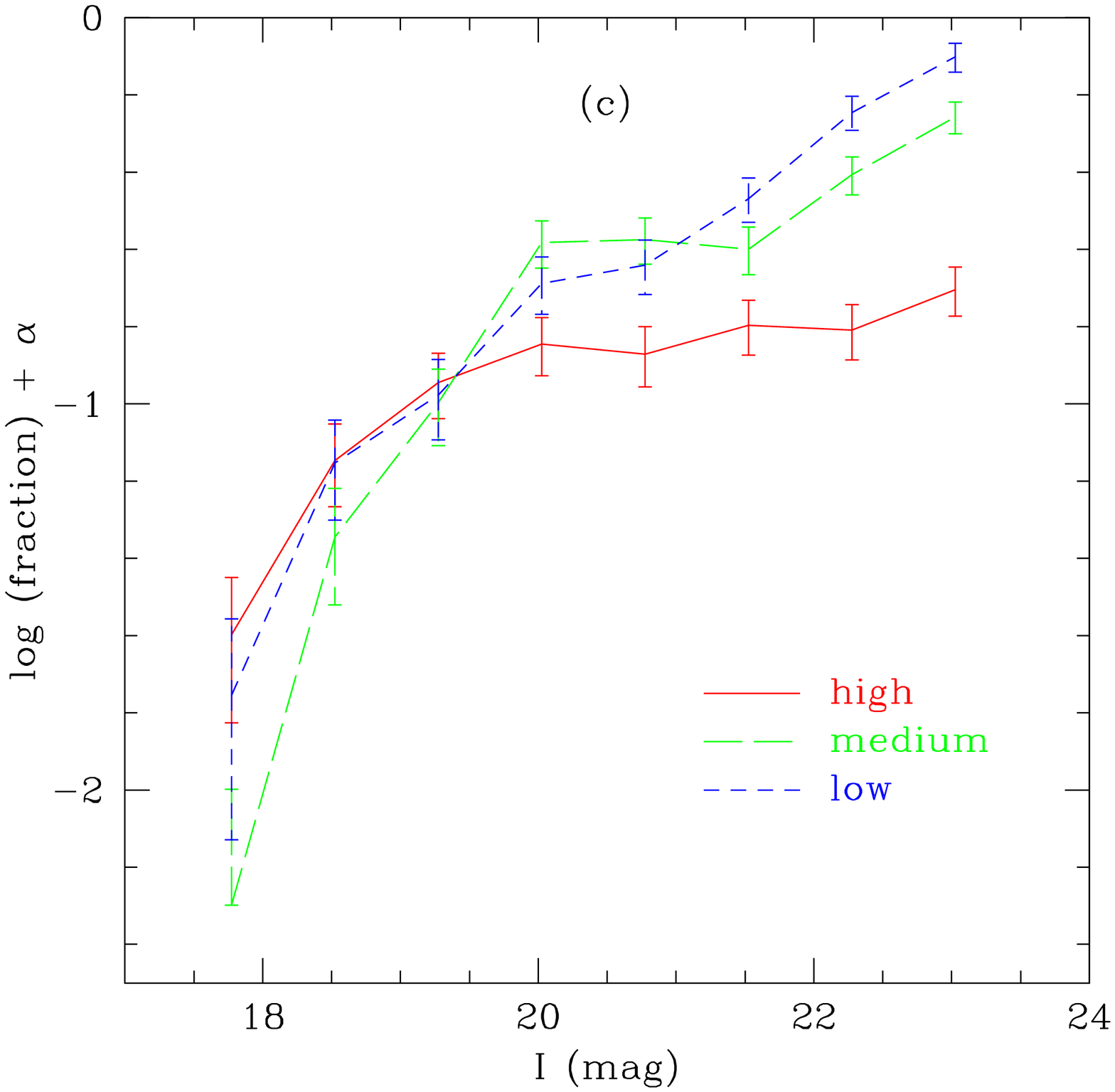,width=2.45in,angle=0}}
\caption{
\scriptsize
(a) The variation in color versus local galaxy density, $\Sigma$,
for cluster members brighter than $I=23.4$.  The open circles and
filled triangles show the galaxies brighter or fainter than $I=21.4$
($M_V^{\ast}$+2), respectively.  The three red lines represent the
loci of the 25, 50, and 75th percentile colors.  The subscript `c'
indicates that the colors are transformed to the equivalent color of an
$I=20.4$ galaxy to take into account the slope of the color-magnitude
relation ($\Delta(V-I)/\Delta I=-0.06$).  The local number density is
calculated from the 10 nearest galaxies and we correct this for residual
field contamination using the blank field data (\S3.2).  (b) The color
distribution of cluster members as a function of local galaxy density.
The lines correspond to the different density regimes: high, medium,
and low shown in Panel (a).  The dotted line gives the distribution for
galaxies in the blank field with photometric redshifts placing them at
$0.32<z_{\rm phot}<0.48$.  We see a trend for typically bluer colors at
lower densities.  (c) The field-corrected luminosity function for cluster
galaxies in the three density slices from Panel (a).  The distributions
are normalized at $I=19.4$ ($M_V^\ast$) to match the high density curve.
A clear steepening of the luminosity function is visible at lower
densities.}
\end{figure*}

\subsection{Environmental dependence of galaxy properties}

By exploiting the striking large scale structure around this cluster, we
can investigate the influence of environment on the photometric properties
of galaxies across more than two orders of magnitude in local density.
We define the environment for each galaxy using the local surface number
density, $\Sigma$, of members, calculated from the 10 nearest neighbors
(including that galaxy) brighter than our magnitude cut, $I=23.4$.
Secondly, we correct for residual field contamination in the redshift
slice.  We statistically subtract galaxies based on the local number
density $\Sigma$ and the color and magnitude distribution of the galaxies
in the blank field (\S2.2) using a Monte Carlo simulation.  The density
$\Sigma$ is also corrected for the field contamination.  The statistical
uncertainty in the resultant color and magnitude distributions of the
cluster members arising from this field correction is negligible.  Even if
we assume a large variation in field density of 20\%, the median color
in the lowest density bin would change only by $\sim$0.02 mag (Fig.~2a).

In the following we use $(V-I)$ color and $I$ magnitude to characterize
the galaxy population. These roughly correspond to $(U-V)$ and $V$ in the
rest-frame, providing a good measure of the relative importance of recent
and past star formation.  We divide the density distribution into three
regimes as indicated in Fig.~2a.  There is a close correspondence between
the local density and structure: the high density region corresponds
to the cluster core within $\lsim 1$\,Mpc; the medium density region
includes the structures defining the filaments surrounding the cluster;
and the low density region comprising the rest of the volume (Fig.~1).

As shown in Fig.~2b, the color distribution in the high density region
is strongly peaked at $(V-I)_c\sim 2$, the color of an early-type cluster
member.  However, as we move to lower  densities, the distribution becomes
dramatically bluer.  In the lowest density regions, there are only a small
fraction (15\%) of galaxies as red as the  early-type color magnitude
sequence seen in higher density regions, with most of the galaxies being
much bluer, $(V-I)_c\sim 1.5$, similar to the field population, which
is dominated by star-forming galaxies at these redshifts (Ellis et al.\
1996).  Fig.~2a shows that this color transition with local density occurs
quite abruptly at $\log_{10}\Sigma\sim 2$, indicating a threshold effect
in transforming galaxy properties.
We performe a $\chi^2$ test to verify the significance of the
sharpness of the color transition, and found that a linear (smooth)
dependence of median color on local density is
rejected at greater than 3 sigma confidence.
The boundary corresponding to this
critical density is highlighted in Fig.~1.  One important clue to the
physical origin of this transition is that over 80\% of the galaxies in
the `transition zone' ($1.8<\log_{10}\Sigma<2.2$) reside in subclumps
outside the core ($>$1.5 Mpc from the cluster center).

The origin of the color-density correlation is hotly debated. It has
been suggested that it may reflect a primordial imprint of the regions
in which the galaxies were formed. However, this is unlikely to give
rise to a sharp transition in the colors of galaxies at a particular
density threshold at the observed epoch. Our data suggest a recent and
environmentally driven transformation that is closely linked to the
galaxy environment at $\log_{10}\Sigma\sim2$.  This sharp color change
is equivalent to supressing the median star formation rate by a factor
of 6, based on the simple `tau' models with exponentially decaying
star formation histories (e.g.\ Kodama \& Bower 2001). Our ability to
pin-point this environment is a fundamental step towards identifying
the dominant mechanism.

There are three mechanisms which are favored in the literature:
ram-pressure stripping (where cold gas in the galaxy disk is removed
as the galaxy passes through a hot intra-cluster medium; Gunn \&
Gott 1972); galaxy--galaxy collisions (which cause cold disk gas to
be driven to the galaxy center creating a star burst; Moore et al.\
1996); and `suffocation' (where {\it warm} gas in the galaxy's halo
is shock heated by the intra-cluster medium so that it can no longer
cool and replenish the cold gas in the disk; Larson et al.\ 1980).
The $\log_{10}\Sigma\sim2$ subclumps are unlikely to have sufficiently
high gas densities (\S3.1), or high enough relative velocities for the
galaxies within them, for ram-pressure stripping to operate effectively
(Abadi et al.\ 1999).  In contrast, we note that the threshold
density corresponds to the point at which the dark matter halos of
individual galaxies will begin to overlap (Brainerd et al.\ 1996).
Under these circumstances, galaxy--galaxy collision become important,
as does `suffocation'. Our data do not distinguish directly between
these possibilities, although they do suggest where to look.
In particular, the two mechanisms predict very different bulge luminosity
functions for the resulting galaxies.
Strong galaxy interactions thicken galaxy disks and brighten bulges,
in contrast, suffocation should leave the
bulge luminosity function unchanged (Dressler 1980; Balogh et al.\ 2001).

It is important to note that the changes in galaxy properties as a
function of local density are most prominent in galaxies fainter than
$\sim 0.1 L_V^\ast$ (Fig.~2a).  This is mirrored to the dramatic change
in the shape of the luminosity function with local galaxy density: where
we see a steepening of the faint end slope of the luminosity function
with decreasing  density (Fig.~2c).  Assuming that their star formation
effectively ceases, the relative absence of low-luminosity galaxies at
projected densities of $\log_{10}\Sigma> 2$ can be understood as they will
fade by $\gsim 1$\,mag as their star formation declines (e.g.\ Kodama \&
Bower 2001).  This will put many of their descendents below our magnitude
limit ($M_V^\ast+4$).  This is also qualitatively consistent with the
presence of large numbers of very low luminosity, $\gg M_V^\ast+3$, passive
dwarf galaxies in local clusters (Binggeli et al.\ 1988; Trentham 1998).

\section{Conclusions}

We have presented a photometric analysis of the galaxy population in
A\,851 at $z=0.41$ using sensitive multiband observations from Subaru.
These cover a  $27'$ (11\,Mpc) field allowing us to compare galaxy
properties in a wide range of environments in the cluster periphery. By
using photometric redshift techniques to identify cluster galaxies
as faint as $M_V^\ast+4$, we have traced the network of filaments and
subclumps around this rich cluster, and been able to investigate the
relation between environment and star formation in faint galaxies.

The colors of faint galaxies show a strong dependence on their local
galaxy density, with an abrupt transition in galaxy colors occurring at
$\log_{10}\Sigma\sim 2$. Thus, for faint galaxies, the transformation of
their properties seems to be almost complete before they have entered
the cluster core.  The changes in the color distribution and in the
luminosity function as a function of local density are coupled: there is
a decreasing fraction of blue, faint galaxies in higher density regions.
This density threshold corresponds to the point at which galaxy halos
lose their individual identity and become incorporated into groups along
the filaments surrounding the cluster.

We have now identified {\it where} the transformation of these galaxies
occurs.  The key remaining question is to identify {\it why}. Two
mechanisms: `suffocation' and galaxy--galaxy collisions are both good
candidates in these environments,
but they can be distinguished by tracing
the variation in the individual components of galaxies, bulges and disks,
between the cluster and the field.

\acknowledgments

We thank the Subaru telescope observatory staff and the Suprime-Cam team
for their invaluable effort to complete the telescope and the instrument
over the last 10 years.  We are grateful to Dr Y.\ Komiyama for his
assistance during our observations.  We acknowledge the Suprime-Cam team
for allowing us to use the blank field data.  We also thank Drs.\ M.\
Balogh, K.\ Pimbblet, N.\ Arimoto, K.\ Shimasaku for helpful discussion.
TK and FN acknowledge the Japan Society for the Promotion of Science for
support through its Research Fellowships for Young Scientists.
IRS acknowledges support from the Royal Society and the Leverhulme Trust.

\end{document}